\documentclass[twocolumn]{aastex61}
\pdfoutput=1 %for arXiv submission
\usepackage{amsmath,amstext}
\usepackage[T1]{fontenc}
\usepackage{apjfonts} 
\usepackage[figure,figure*]{hypcap}
\usepackage{verbatim}
\shorttitle{LIERs at $z\sim0.9$}
\shortauthors{Hviding et al.}

\begin{document}

\title[GOODS LIERs]{Spatially Extended Low Ionization Emission Regions (LIERs) at $\lowercase{z}\sim0.9$}

\author{Raphael E. Hviding}
\affiliation{Department of Physics and Astronomy, Dartmouth College, Hanover, NH 03755, U.S.A.}
\affiliation{Department of Astronomy, Steward Observatory, University of Arizona, 933 North Cherry Avenue, Rm N204, Tucson, AZ, 85721, USA}
\author{Gabriel B. Brammer}
\affiliation{Space Telescope Science Institute, 3700 San Martin Drive, Baltimore, MD 21218, U.S.A.}
\affiliation{Cosmic Dawn Center, Niels Bohr Institute, University of Copenhagen, Juliane Maries Vej 30, DK-2100 Copenhagen \O, Denmark}
\author{Ivelina G. Momcheva}
\affiliation{Space Telescope Science Institute, 3700 San Martin Drive, Baltimore, MD 21218, U.S.A.}
\author{Britt Fisher Lundgren} 
\affiliation{Department of Physics, University of North Carolina Asheville, One University Heights, Asheville, NC 28804, U.S.A.}
\author{Danilo Marchesini} 
\affiliation{Department of Physics and Astronomy, Tufts University, 574 Boston Avenue Suites 304, Medford, MA 02155, U.S.A.}
\author{Norbert Pirzkal}
\affiliation{Space Telescope Science Institute, 3700 San Martin Drive, Baltimore, MD 21218, U.S.A.}
\author{Russell E. Ryan}
\affiliation{Space Telescope Science Institute, 3700 San Martin Drive, Baltimore, MD 21218, U.S.A.}
\author{Andrea Vang}
\affiliation{Department of Astronomy, University of Wisconsin-Madison, 475 N. Charter St., Madison, WI 53706, U.S.A.}
\author{David A. Wake}
\affiliation{Department of Physics, University of North Carolina Asheville, One University Heights, Asheville, NC 28804, U.S.A.}
\author{Matthew Bourque}
\affiliation{Space Telescope Science Institute, 3700 San Martin Drive, Baltimore, MD 21218, U.S.A.}
\author{Catherine Martlin}
\affiliation{Space Telescope Science Institute, 3700 San Martin Drive, Baltimore, MD 21218, U.S.A.}
\author{Kalina V. Nedkova}
\affiliation{Department of Physics and Astronomy, Tufts University, 574 Boston Avenue Suites 304, Medford, MA 02155, U.S.A.}

\begin{abstract}
We present spatially resolved emission diagnostics for eight $z\sim0.9$ galaxies that demonstrate extended low ionization emission-line regions (LIERs) over kpc scales. Eight candidates are selected based on their spatial extent and emission line fluxes from slitless spectroscopic observations with the HST/WFC3 G141 and G800L grisms in the well-studied GOODS survey fields. Five of the candidates (62.5\%) are matched to X-ray counterparts in the \textit{Chandra X-Ray Observatory} Deep Fields. We modify the traditional Baldwin-Philips-Terlevich (BPT) emission line diagnostic diagram to use [SII]/(H$\alpha$+[NII]) instead of [NII]/H$\alpha$ to overcome the blending of [NII] and H$\alpha$+[NII] in the low resolution slitless grism spectra. We construct emission line ratio maps and place the individual pixels in the modified BPT. The extended LINER-like emission present in all of our candidates, coupled with X-Ray properties consistent with star-forming galaxies and weak [OIII]$\lambda$5007\AA\ detections, is inconsistent with purely nuclear sources (LINERs) driven by active galactic nuclei. While recent ground-based integral field unit spectroscopic surveys have revealed significant evidence for diffuse LINER-like emission in galaxies within the local universe $(z\sim0.04)$, this work provides the first evidence for the non-AGN origin of LINER-like emission out to high redshifts.
\end{abstract}


\section{Introduction}
Introduced in \citet{Heckman1980LINER} as a new class of galactic nuclei, low ionization nuclear emission-line regions (LINERs) exhibit emission-line ratios consistent with low-ionization atomic transitions, with strong [NII], [SII], and [OI] and relatively weak [OIII]. Originally, LINERs were hypothesized to form part of a continuum of active galactic nuclei (AGNs) ranging from high luminosity quasars and Seyfert galaxies towards their low luminosity counterparts. In recent years, LINERs have been of particular interest due to the mounting evidence that their characteristic emission is not constrained solely to the nucleus, but extends out to kpc scales as shown by several integral field unit (IFU) surveys \citep[e.g.][]{Sarzi2006sauron1, Sarzi2010sauron2, Singh2013califa, Belfiore2016manga1}.

Although LINERs have been identified in many galaxies using a variety of spectral apertures, we adopt the \citep{Belfiore2016manga1} naming convention, which excludes the 'N' for nuclear and refers to low-ionization emission line region as LIERs. In the local universe, up to 40\% of `normal' galaxies (excluding interacting galaxies, merging galaxies, and active galaxies) may be LIERs in certain stellar mass ranges \citep{Belfiore2016manga1}. 

The literature surrounding diffuse LIERs has advanced several hypotheses for the extended emission. Firstly, the `stellar hypothesis' for the extended emission \citep[e.g.][]{Binette1994PAGB, Stasiska2008PAGB, Fernandes2011PAGB, Yan2012PAGB}  argues that post asymptotic giant branch (post-AGB) stars provide sufficient photoionization to produce observed LIER ratios, a compelling theory as LINER-like emission is predominantly observed in early-type and red galaxies, where evolved stars become the dominant ionizing source \citep{Belfiore2016manga1}. Secondly, the 'starburst hypothesis' for the extended emission \citep{Sharp10GalWinds,Yuan10StarburstAGN,Ho14StarburstWinds,Ho16GalWinds} argues that a burst of star-formation, leading to ionizing outflows and galactic winds, causing the elevated line ratios on extended scales. Thirdly, the 'merger hypothesis' for extended emission \citep[e.g.][]{Monreal06LINERMerger, Monreal10TidalShock, Soto10MergerShocks, Rich2011MergerShocks} argues that tidally induced gas flows from galaxy interaction leading to shock ionization across various spatial extents. Finally, recently \citet{Weilbacher18Antennae} argued that Lyman-continuum leakage was sufficient to explain the photoionization present in the merger Antennae Galaxy.

The purpose of this work is to extend evidence for the existence of LIERs out to $z\sim0.9$ and present methods for their detection using slitless grism spectroscopy. So far spatially resolved spectroscopic observations of LIERs galaxies have been limited to the local universe. The average redshift of the LIERs confirmed by \citet{Belfiore2016manga1} in the Sloan Digital Sky Survey-IV (SDSS-IV) Mapping Nearby Galaxies at Apache Point Observatory \citep[MaNGA]{Bundy2015MaNGA} survey is $\langle z \rangle \sim 0.04$. We report for the first time the detection of eight galaxies at $z\sim0.9$ with spatially-extended LINER-like emission line ratios (i.e., LIERs).  In order to detect spatially resolved emission diagnostics of $z\sim0.9$ galaxies, we make use of the 3D-HST survey \citep{Brammer123DHST, Momcheva163dhst}, a near-infrared slitless grism survey covering most of the Cosmic Assembly Near-infrared Deep Extragalactic Legacy Survey (CANDELS) Treasury survey area. The 3D-HST Survey carried out two-orbit depth slitless spectroscopy with the G141 grism (1.10 to 1.65 $\mu$m) on the {\it Hubble Space Telescope} ({\it HST}) Wide Field Camera 3 (WFC3) instrument. Additional {\it HST}/WFC3 G102 grism spectroscopy (0.8 to 1.1 $\mu$m) is available from the following HST programs: GO-11359 (PI: O'Connell), GO-12190 (PI: Koekemoer), GO-13420 (PI: Barro), GO-13779 (PI: Malhotra), GO-14227 (PI: Papovich). We focus on the Great Observatories Origins Deep Survey (GOODS) fields which have both the deepest slitless spectra over a wide area and extensive deep X-ray coverage -- 7 Msec in CDFS \citep{Luo17CDFS} and 2 Msec in CDFN \citep{Xue16CDFN} -- allowing for the use of additional X-ray AGN selection criteria.  

The discovery of LIERs at $z\sim0.9$ is greatly aided by the development of \texttt{Grizli}\footnote{https://github.com/gbrammer/grizli}, a new library for the analysis of space-based slitless spectroscopy (Brammer et al., in prep.). The \texttt{Grizli} library presents two significant improvements to the \citet{Brammer123DHST} and \citet{Momcheva163dhst} 3D-HST reductions. First, \texttt{Grizli} allows for simultaneous fitting of all available exposures from the G102 and G141 grisms, combining grism spectra obtained at multiple position angles. And second, \texttt{Grizli} provides a mechanism to create drizzled emission line maps, again from combining all available grisms and position angles. We use these emission line maps to create the emission line ratio maps used in this work. 

This paper is structured as follows. In Section~\ref{sec:LineBlend} we describe our modification of traditional LINER diagnostics in order to account for the line-blending of low-resolution grism spectroscopy. Section~\ref{sec:Selection} describes the selection of candidates and their extraction using the new \texttt{Grizli} pipeline. Section~\ref{sec:LIER} presents the candidates in detail, including spatially resolved spectroscopic diagnostics that demonstrate signatures consistent with spatially extended LIERs. Section~\ref{sec:Conclusions} summarizes our findings and discusses future research. 

Cosmological calculations are made assuming the following $\Lambda$CDM model from WMAP measurements with $H_0 = 69.3\,$km$\,$s$^{-1}\,$Mpc$^{-1}$, $\Omega_{M} = 0.287$, and $\Omega_{\Lambda} = 0.713$ \citep{Hinshaw13WMAP}. 

\begin{figure*}
\centering
	\includegraphics[width=2.09\columnwidth]{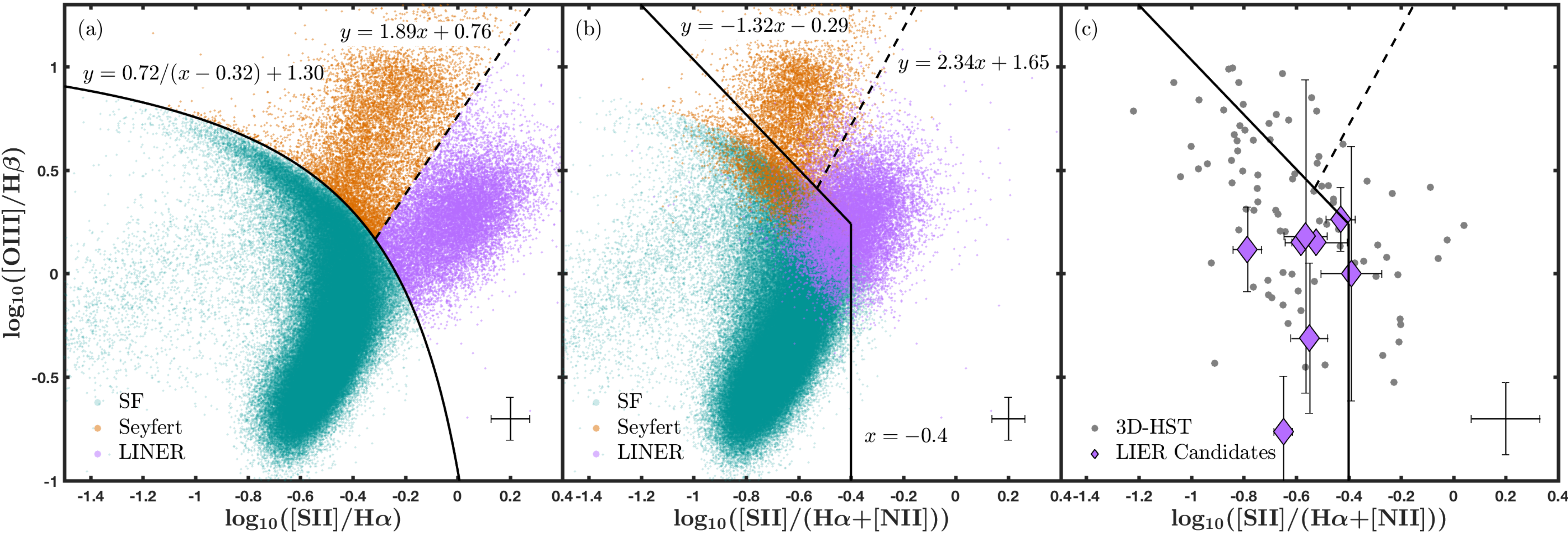}
	\caption{BPT excitation diagnostic diagrams. Panels (a) and (b) plot the \citet{Oh2011SDSS} Sloan Digital Sky Survey Data Release 7 emission line fluxes on a traditional BPT and line-blended BPT respectively. In panel (a) we plot the \citet{Kewley2006BPT} delineations which inform our modified BPT delineations, shown in panels (b) and (c). We plot integrated GOODS 3D-HST line measurements for objects with a 3$\sigma$ detection in each line and the integrated \texttt{Grizli} line measurements for our LIER candidates in panel (c). Median errors are shown in each panel.}
	\label{fig:BPTLDApaper}  
\end{figure*}

\section{Line Blending in 3D-HST}
\label{sec:LineBlend}

In order to categorize the emission line ratios of the selected galaxies, we make use of the well known BPT diagnostic \citep{Baldwin1981BPT}. In order to distinguish between star forming galaxies and typical LINERs, we employ the BPT diagram that compares the [SII]$\lambda$6717\AA+6731\AA\ to H$\alpha\lambda$6563\AA\ ratio to the [OIII]$\lambda$5007\AA\ to H$\beta\lambda$4861\AA\ ratio.\footnote{For convenience, emission lines shall be referenced by species alone.} Between the redshifts of $z\sim0.7$ and $z\sim1.5$ the H$\alpha$ and [SII] emission lines falls into the wavelength coverage of the G141 grism, while the H$\beta$ and [OIII] lines fall in the coverage of the G102 grism. However, due to the low spectral resolution ($R\sim 150$) of the {\it HST}/WFC3 G141 grism, the entire [NII]$\lambda$6548\AA, H$\alpha\lambda$6563\AA, [NII]$\lambda$6584\AA, complex\footnote{[NII] shall reference the [NII]$\lambda$6584+6584\AA\ complex.} of line is not resolved, but blended into a single line \citep{Brammer123DHST}. 

We develop a modification to the traditional BPT diagnostic in order to account for this line-blending in H$\alpha$. We use the \citet{Oh2011SDSS} improved emission line measurements from the \citet{Abazajian09DR7} SDSS Data Release 7 (DR7) catalog in order to create a modified BPT diagnostic to select all objects with a signal to noise ratio (SNR) $>3$ in each aforementioned line. In Figure \ref{fig:BPTLDApaper}(a) we show all selected objects color-coded using the \citet{Kewley2006BPT} criteria to separate star forming galaxies, Seyferts, and LINERs. We then plot the objects after adding in the [NII] contribution into the H$\alpha$ line as shown in Figure \ref{fig:BPTLDApaper}(b). This is our modified BPT diagram. In order to mimic the original selection criteria, we generate two classifications in this modified BPT diagram. To separate Seyferts and LINERs we use a two class linear discriminant analysis \citep[LDA]{LDA} to find the linear subspace which minimizes the within-class scatter and maximizes the across-class scatter. After projecting onto the optimal linear subspace, we choose a cutoff between classes when the fraction of correctly identified LINERs is equal to that of Seyferts (approximately 88\%). This results in Equation \ref{eq:AGN}, \begin{equation}
\label{eq:AGN}
\log_{10}\left(\frac{[\textrm{OIII}]}{\textrm{H}\beta}\right) = 2.34\cdot\log_{10}\left(\frac{[\textrm{SII}]}{\textrm{H}\alpha+[\textrm{NII}]}\right) + 1.65
\end{equation} which provides a distinction between LINERs and Seyferts in the modified BPT. We make use of a linear classifier due to it's simplicity and since a higher order function would not drastically reduce the inter-class contamination.

We conduct the same procedure to generate a selection cut between all AGN and star forming galaxies. Again, we generate a two class LDA but instead select our cutoff in order to purely select AGN, without any galaxies lying above the line, resulting in Equation \ref{eq:Gal}. \begin{equation}
\label{eq:Gal}
\log_{10}\left(\frac{[\textrm{OIII}]}{\textrm{H}\beta}\right) = -1.32\cdot\log_{10}\left(\frac{[\textrm{SII}]}{\textrm{H}\alpha+[\textrm{NII}]}\right) - 0.29
\end{equation} This results in only correctly identifying 57\% of all AGN, a relatively conservative criterion in order to ensure that there is minimal star forming galaxy contamination. In addition, we add in a vertical component to the criterion, categorizing any object with a $\log_{10}([\textrm{SII}]/([\textrm{H}\alpha]+[\textrm{NII}])) > -0.4$ as an AGN. This was assigned visually and only adds less than 4\% contamination to the total AGN sample. We make use of the latter criterion extensively in Section \ref{sec:LIER} where many objects do not have corresponding [OIII]/H$\beta$ ratios and we solely rely on the [SII]/(H$\alpha+$[NII]) ratio. 

In Figure \ref{fig:BPTLDApaper}(c) we plot all 3D-HST extracted objects in the GOODS fields that have a 3$\sigma$ detection in all of the contributing lines, 86 in total, for comparison. 

\section{Sample Selection and Properties}
\label{sec:Selection}

\subsection{Emission Line Flux}
\label{sec:fluxcuts}
We begin by selecting objects from the \citet{Momcheva163dhst} catalog, which presents reduced data and data products from WFC3 G141 grism spectroscopy from the 3D-HST survey, a 248-orbit HST Treasury program in the CANDELS fields. We use the emission line catalog, including photometry correction scales, to find candidates with high signal to noise emission line ratios. We impose a flux limit on both the H$\alpha$+[NII] and [SII] emission lines, requiring a minimum H$\alpha$+[NII] flux of $30\times10^{-17}\textrm{erg}\,\textrm{cm}^{-2}\,\textrm{s}^{-1}$ and a minimum [SII] flux of $8\times10^{-17}\textrm{erg}\,\textrm{cm}^{-2}\,\textrm{s}^{-1}$, which let us consistently detect H$\alpha$+[NII] emission out to 9 kpc and [NII] emission out to 6 kpc in most objects. Following this selection, we are left with 795 sources in the GOODS-South field and 687 in the GOODS-North field. In order to retrieve well constrained emission line ratios with low uncertainties, we require candidates to have an H$\alpha$+[NII] emission SNR $>$ 10, and an [SII] emission SNR $>$ 2,  resulting in 40 objects in the GOODS-South field and 61 in the GOODS-North field. 

\subsection{Source Geometry for Optimal Line Map}
\label{sec:geometry}

We further select objects where the combination of the morphology and the geometry of the dispersed spectra is favorable to determining robust emission line ratio maps. Due to the low resolution of the grism spectra, the  $\sim275$\AA\ (observed frame) separation between H$\alpha$ and the [SII] doublet results in a separation between the two features of only $\sim6$ pixels in the images. The \texttt{Grizli} pipeline does mask areas where there is contamination from neighboring lines but then we can not constrain emission line ratios in these regions. This effect is only significant in the dispersion direction of the image and is exacerbated by the morphology of the galaxy. 

\begin{table*}\centering
\tabcolsep=0.1cm
\caption{LIER Candidates}
\label{tab:LIERCands}
\begin{tabular}{cccccccrccccccc}
\hline
\hline
  \multicolumn{1}{c}{ID} &
  \multicolumn{1}{c}{RA} &
  \multicolumn{1}{c}{Dec} &
  \multicolumn{2}{c}{$z$} &
  \multicolumn{1}{c}{$J_H$} &
  \multicolumn{1}{c}{M${_\star}^\textrm{a}$} &
  \multicolumn{1}{c}{sSFR$^\textrm{a}$} &
  \multicolumn{1}{c}{${L_{\text{0.5-7keV}}}^\textrm{a,b}$} &
  \multicolumn{1}{c}{ID} &
  \multicolumn{1}{c}{${L_{\text{0.5-7keV}}}^\textrm{a}$} &
  \multicolumn{1}{c}{$\Gamma_\text{eff}$} &
  \multicolumn{1}{c}{$\underline{f_\textrm{X}}^\textrm{a}$} &
  \multicolumn{1}{c}{$\underline{L_{\text{0.5-7keV}}}^\textrm{a}$} &
  \multicolumn{1}{c}{$\underline{f_\textrm{X}}^\textrm{a}$} \\
  \multicolumn{1}{c}{3D-HST} &
  \multicolumn{2}{c}{J2000} &
  \multicolumn{1}{c}{{ 3D-HST$^\textrm{c}$}} &
  \multicolumn{1}{c}{{\texttt{Grizli}}} &
  \multicolumn{1}{c}{AB} &
  \multicolumn{1}{c}{M$_\odot$} &
  \multicolumn{1}{c}{yr$^{-1}$} &
  \multicolumn{1}{c}{erg\,s$^{-1}$} &
  \multicolumn{1}{c}{CDF} &
  \multicolumn{1}{c}{erg\,s$^{-1}$} &
  \multicolumn{1}{c}{} &
  \multicolumn{1}{c}{$f_R$} &
  \multicolumn{1}{c}{$L_{\text{1.4GHz}}$} &
  \multicolumn{1}{c}{$f_{K_s}$} \\
\hline
  \multicolumn{13}{c}{GOODS$-$S} \\
\hline
   983  & 03$^\textrm{h}$32$^\textrm{m}$29$^\textrm{s}$.44 & $-$27$^\circ$55$'$38$''$.20 & 0.660 & 0.660 & 19.84 & 10.27 &  $-$8.66 & 41.4 & --- &  ---  &  --- &   ---   & --- &   ---  \\
  15738 & 03$^\textrm{h}$32$^\textrm{m}$25$^\textrm{s}$.86 & $-$27$^\circ$50$'$19$''$.71 & 1.095 & 1.095 & 21.04 & 10.31 &  $-$8.65 & 41.6 & 466 & 41.38 & 1.95 & $-$2.55 & --- & $-$3.03\\
  26087 & 03$^\textrm{h}$32$^\textrm{m}$39$^\textrm{s}$.88 & $-$27$^\circ$47$'$15$''$.04 & 1.097 & 1.095 & 20.90 & 10.33 &  $-$8.66 & 41.6 & 749 & 41.52 & 2.52 &   ---   & --- &   ---  \\
  36653 & 03$^\textrm{h}$32$^\textrm{m}$36$^\textrm{s}$.03 & $-$27$^\circ$44$'$23$''$.76 & 1.038 & 1.044 & 21.55 & 10.08 &  $-$8.64 & 41.3 & 677 & 41.66 & 1.43 & $-$2.10 & --- & $-$2.58\\
\hline
  \multicolumn{13}{c}{GOODS$-$N} \\
\hline
  18043 & 12$^\textrm{h}$37$^\textrm{m}$22$^\textrm{s}$.57 & $+$62$^\circ$13$'$56$''$.74 & 1.022 & 1.023 & 20.81 & 10.49 & $-$8.72 & 41.7 & --- &  ---  & ---  &  ---    &  ---  &   ---  \\
  21285 & 12$^\textrm{h}$36$^\textrm{m}$59$^\textrm{s}$.92 & $+$62$^\circ$14$'$49$''$.96 & 0.761 & 0.761 & 20.50 & 10.31 & $-$8.69 & 41.5 & 416 & 41.40 & 1.40 & $-$2.31 & 18.08 & $-$2.79\\
  22593 & 12$^\textrm{h}$37$^\textrm{m}$01$^\textrm{s}$.94 & $+$62$^\circ$15$'$10$''$.43 & 0.938 & 0.946 & 21.42 & 10.04 & $-$8.84 & 41.1 & --- &  ---  & ---  &  ---    &  ---  &   ---  \\
  22815 & 12$^\textrm{h}$37$^\textrm{m}$08$^\textrm{s}$.37 & $+$62$^\circ$15$'$14$''$.66 & 0.839 & 0.842 & 20.98 & 10.02 & $-$8.62 & 41.3 & 471 & 41.79 & 1.40 & $-$2.04 &  ---  & $-$2.51\\
\hline  
\end{tabular}
\raggedright{$^\text{a}$\, Values quoted are the log$_{10}$ of the measurement.\\
$^\text{b}$\, Predicted X-ray luminosity derived from the star formation rates and stellar masses following \citet{Lehmer16BHB}.}\\
$^\text{c}$\, Spectroscopic.

\end{table*}
\begin{figure*}
\centering
	\includegraphics[width=2.09\columnwidth]{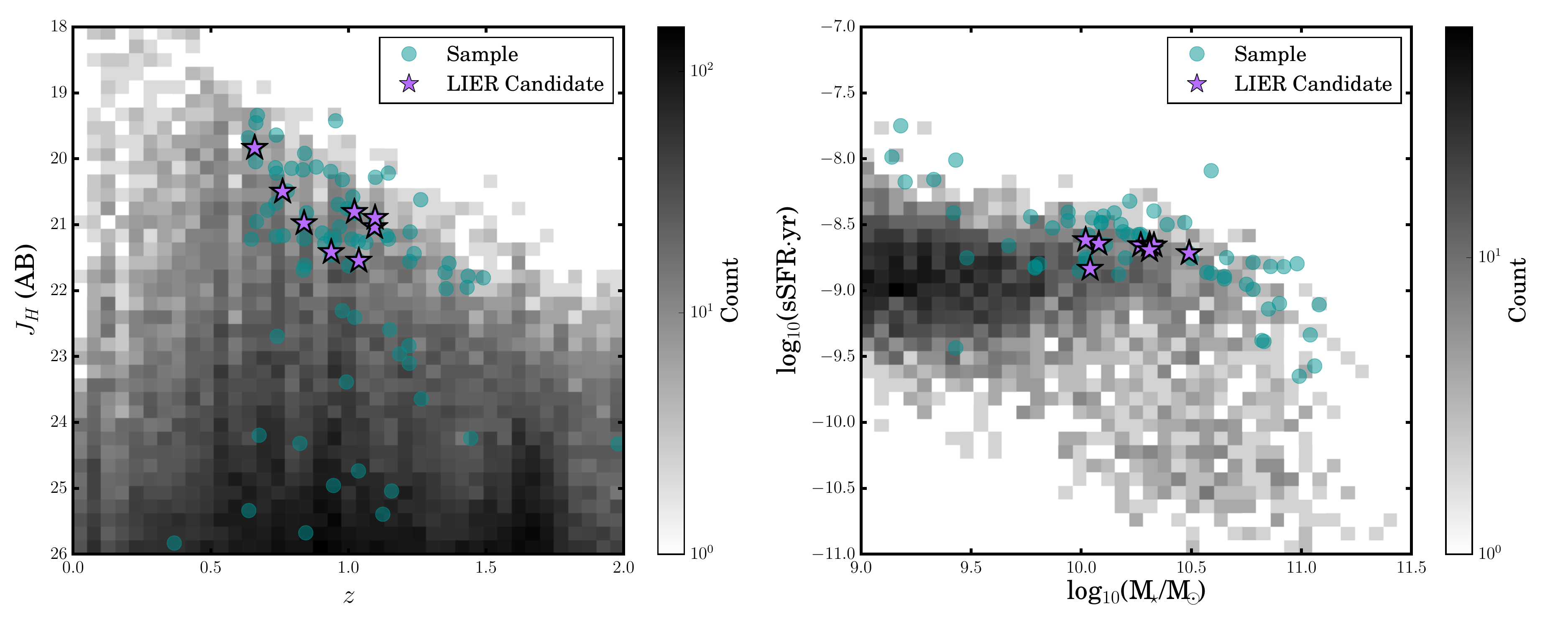}
	\caption{LIER candidates plotted as stars relative to all GOODS 3D-HST objects from \citet{Skelton143DHST} and objects satisfying our flux, SNR and morphological selection cuts. In (a) the candidates are shown with respect to their $J_H$ band magnitude against their redshift, while in (b) only objects with  $0.5 < z < 1.5$ and $\log_{10}(\textrm{M}_\star/\textrm{M}_\odot) > 9$ are shown with their specific star formation rates plotted against their total stellar mass.}
	\label{fig:HISTpaper}  
\end{figure*}

In order to select galaxies with the least contamination, we use the structural parameters measured for 3D-HST \citep{vdWal143dhstgalfit} to select candidates where the grism spectral dispersion axis is not aligned to the morphological position angle. Round objects (i.e., with an axis ratio $q = (a/b) \sim 1$) are unaffected by the relative directions of the dispersion and position angles. We therefore only select objects with a $q < 0.8$ in order to enhance the separation of the emission lines, reducing our sample to 33 sources in the GOODS-South field and 50 in the GOODS-North field. Finally we impose a condition on the alignment between the dispersion axis and the morphological position angle, requiring at least a $10^\circ$ separation between the two. Together, our morphological and spatial cuts ensure that our H$\alpha$ and [SII] emission are separated by at least $6$ pixels. For objects with multiple grism angles we use the angle that maximizes the separation parameter. Our final sample is comprised of 67 sources, 28 in the GOODS South field and 39 in the GOODS North field. 

\subsection{Final Sample}
\label{sec:finalsamp}
For all objects satisfying the flux, SNR, morphology and position angle cuts, we extract spatially resolved emission line maps for each line in the modified BPT diagnostic using \texttt{Grizli}. We further divide the [SII] map by the H$\alpha$+[NII] map and the [OIII] map by the H$\beta$ map to create the emission line ratio maps discussed in Section \ref{sec:LIER}. Given the few number of objects, we examined all ratio maps individually by eye and determined those with high visual evidence for extended LINER-like emission if there was consistent detection of elevated line ratios, i.e. similar adjacent pixels, out to kpc scales. The final sample we selected consists of the eight objects listed in Table \ref{tab:LIERCands} alongside the \texttt{Grizli} derived redshifts, J-band magnitude, stellar mass \citep[determined from SEDs and space-based infrared photometry]{Skelton143DHST}, star formation rate (UV+IR based, see \citet{Whitaker14SFR} for details), and X-Ray data \citep{Luo17CDFS,Xue16CDFN}. 

\begin{figure*}
\centering
	\includegraphics[width=2.09\columnwidth]{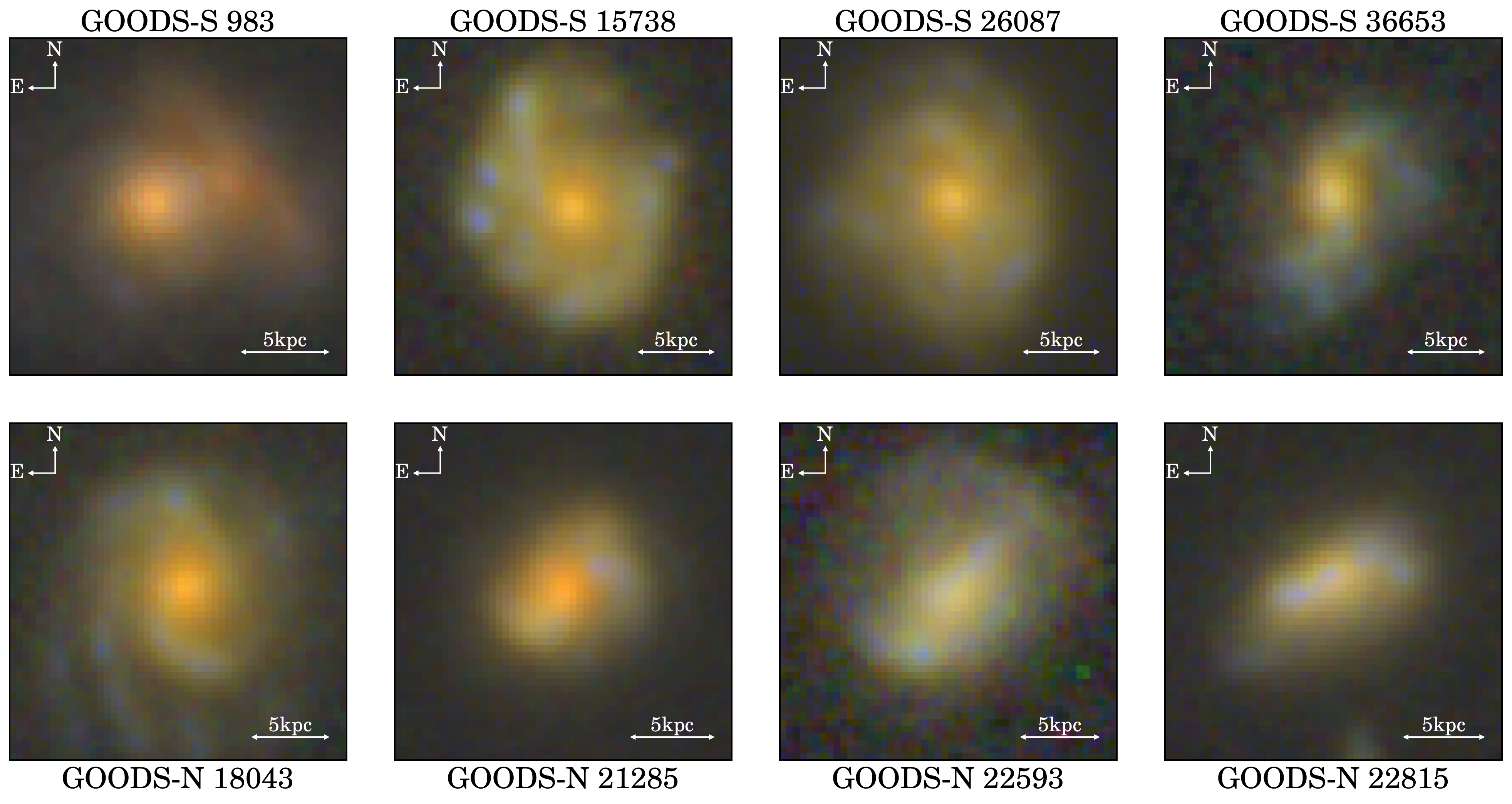}
	\caption{Three-color images of our LIER candidates with the following channels: R = F160W, G = F125W, B = F775W, all convolved to the F160W point spread function. Each image is $3''\times3''$ aligned with north up and east left. The color combination was done using the method of \citet{Lupton04RGB}.}
	\label{fig:RGBpaper}  
\end{figure*}
\begin{figure*}
\centering
	\includegraphics[width=2.09\columnwidth]{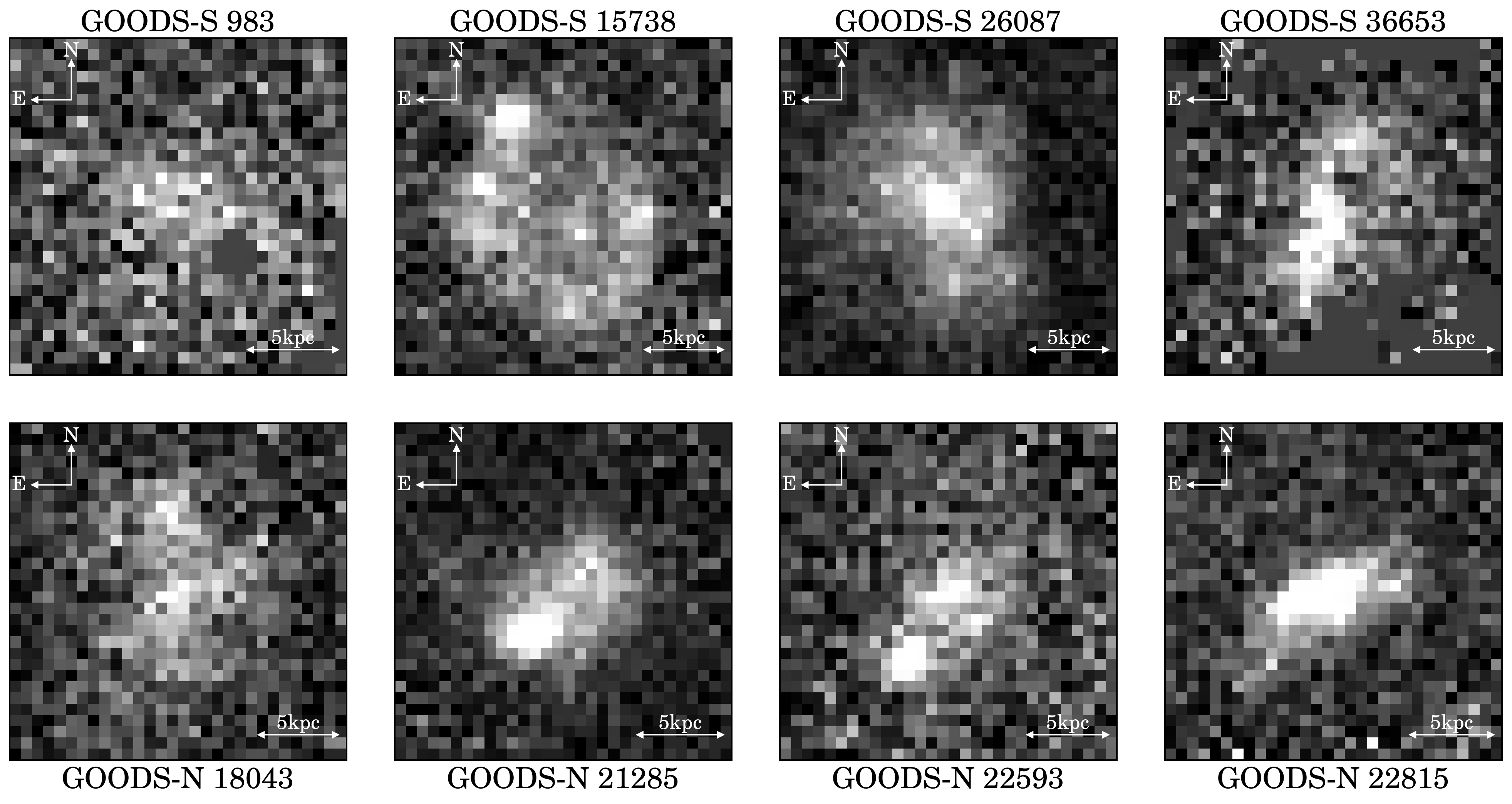}
	\caption{Spatially resolved H$\alpha$ + [NII] emission maps for each of our LIER candidates extracted using the 
	\texttt{Grizli} pipeline. Each image is $3''\times3''$ aligned with north up and east left. 
}
	\label{fig:HaMAPpaper}  
\end{figure*}

Figures \ref{fig:BPTLDApaper}(c) and \ref{fig:HISTpaper} show our final sample with respect to the full 3D-HST catalog. Figure \ref{fig:HISTpaper} additionally plots objects satisfying our flux, SNR, and morphological selection criteria, `i.e.' cuts corresponding to physical properties. Figure \ref{fig:BPTLDApaper}(c) shows the integrated emission line ratios of the modified BPT diagram with our final sample of eight object shown with large symbols. Most fall within a region that, while denoted as star forming with our modified diagnostic, contains significant overlap with LINERs, and may potentially be identified as LINER AGN if followed up with high resolution spectroscopy not limited by H$\alpha$ line blending. 

Figure \ref{fig:HISTpaper}(a) plots our objects according to their \texttt{Grizli} redshifts and 3D-HST $J_H$ magnitudes. Our candidates occupy a fairly narrow redshift range ($0.66-1.52$), a result of our requirement that the emission features fall within the wavelength coverage of the grisms. Furthermore, at any given redshift, our objects are located on the brighter end of the magnitude distribution due to our flux constraints on the candidates, along with the majority of our selected objects. Figure \ref{fig:HISTpaper}(b) shows our candidates relative to objects in a similar redshift range $(0.5 < z < 1.5)$ and mass range $(\log_{10}(\textrm{M}_\star/\textrm{M}_\odot) > 9)$, plotting their total stellar mass against their specific star formation rate. While the sample objects are distributed fairly similarly to the entire sample over specific star formation rates, they are constrained to the high end of the stellar mass distribution, between $10^{10.0}$ and $10^{10.5}$ M$_\odot$, again likely attributed to our flux cuts. We find that our LIER candidates do not differ from the set of all objects satisfying our selection criteria, as demonstrated by Figure \ref{fig:HISTpaper}, reinforcing that the sample properties are dominated by our cuts. Compared to the local sample presented by \citet{Belfiore2016manga1}, the  candidates occupy a stellar mass range where, in the local universe, 20\% of galaxies, excluding merging, interacting, or active galaxies, may be LIERs. 

\begin{figure*}
\centering
	\includegraphics[width=2.09\columnwidth]{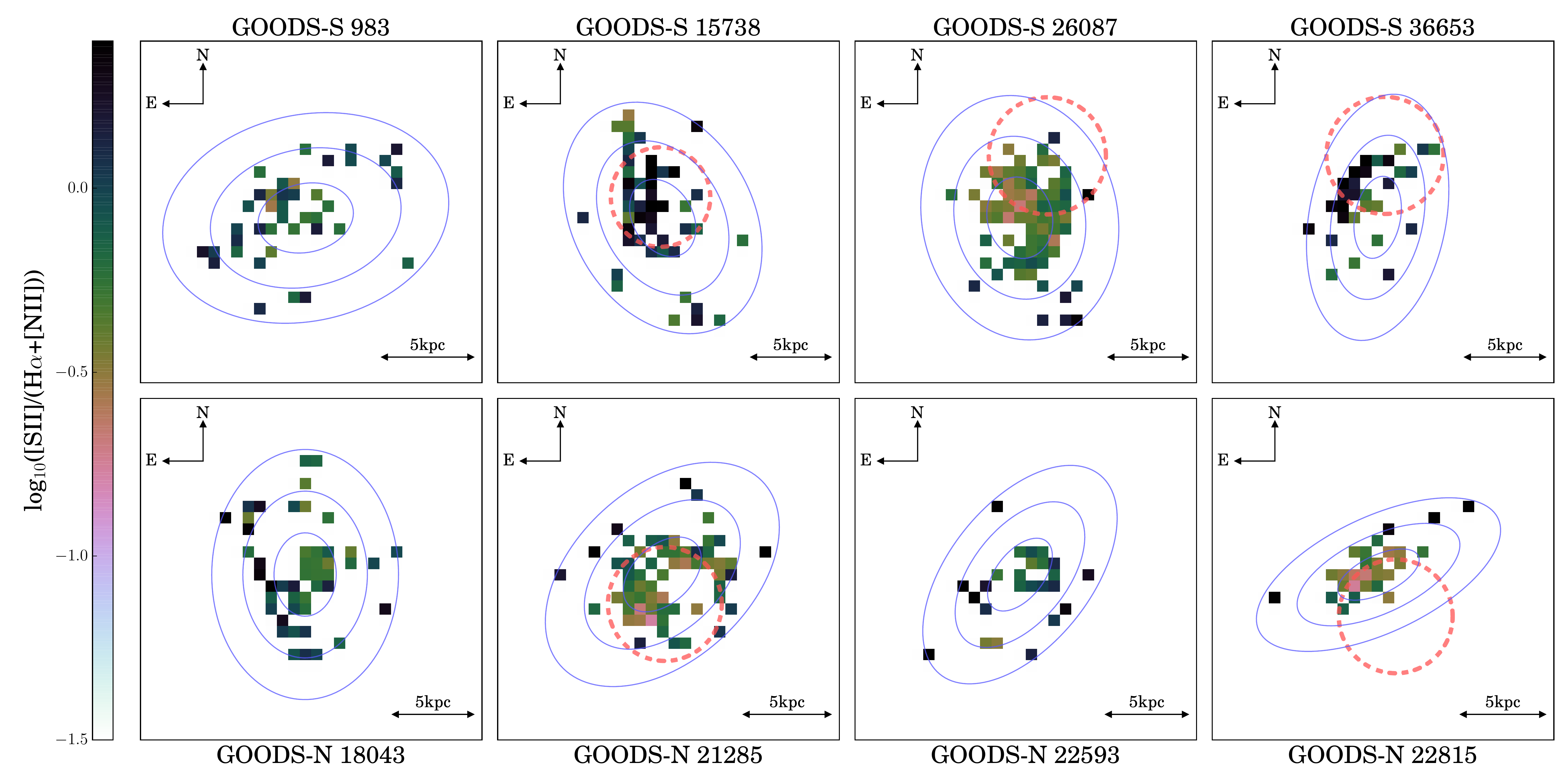}
	\caption{Spatially resolved [SII]/(H$\alpha$ + [NII]) emission maps for each of our LIER candidates with the proper galactic distances of 3, 6, and 9 kpc denoted with green rings. Pixels are shown only if the [SII] flux has a 2$\sigma$ detection. The lower limit of the ratio is shown instead if there is a $<1\sigma$ detection in H$\alpha$ + [NII].  For the objects with a matched \textit{Chandra} detection, the 1$\sigma$ X-ray position is shown as a dashed red circle. Each image is $3''\times3''$ aligned with north up and east left. The log of the emission line ratio is mapped to the perceptually uniform and colorblind friendly colormap outlined in \citet{Green11Cubehelix} known as `cubehelix'.
}\label{fig:LINERpaper}  
\end{figure*}

We match the sample of eight candidates against the X-ray catalogs in the two fields \citep{Luo17CDFS,Xue16CDFN} with a match tolerance of three arcseconds. All our objects fall within the \textit{Chandra} footprint. Five of our candidates are matched with an X-ray source: GOODS-S 15738, GOODS-S 26087, GOODS-S 36653, GOODS-N 21285, GOODS-N 22815. Objects are classified as AGN following the criteria outlined in \citet{Luo17CDFS} as follows: (1) $L_\textrm{X,int} \ge 3\times10^{42} \, \textrm{erg}\,\textrm{s}^{-1}$ (luminous X-ray sources), (2) $\Gamma_\textrm{eff} \le 1.0$ (hard X-ray sources), (3) $\log(f_\textrm{X}/f_R)>-1$ ($f_\textrm{X}$ is the, in order of priority, the full-band, soft-band, or hard-band detected flux; $f_R$ is the $R$-band flux), (4) $L_\textrm{X,int}/L_\textrm{1.4GHz} \ge 2.4\times10^{18}$, (5) $\log(f_\textrm{X}/f_{K_s}) > -1.2$ ($f_{K_s}$ is the $K_s$-band flux). The X-ray luminosity, $\Gamma_\textrm{eff}$, $\log(f_\textrm{X}/f_R)$, $L_\textrm{X,int}/L_\textrm{1.4GHz}$, and $\log(f_\textrm{X}/f_{K_s})$ for the five X-ray detected sources are listed in Table \ref{tab:LIERCands}. None of the five candidates satisfy any of the applicable \citet{Luo17CDFS} AGN criteria, i.e., in all five the X-ray flux is likely due to stellar origins.

In addition, we estimate the expected X-ray luminosity due to low and high mass black hole binaries to compare to the \textit{Chandra} observations. Our estimates don't include the contribution due to hot gas, but still likely capture the majority of the luminosity, given that galactic X-ray flux above 1.5 keV is dominated by black hole binaries. \citep{Hornschemeier05X,Fragos13XB,Lehmer10X,Lehmer16BHB}. We use the \citet{Lehmer16BHB} scaling relationships given by: 
\begin{align}\label{equ:xrb}
\text{L}_{x,LMXB}(z)&=\alpha_0(1+z)^\gamma M_* ,\nonumber \\
\text{L}_{x,HMXB}(z)&=\beta_0(1+z)^\delta SFR ,
\end{align}
where $\log_{10}(\alpha_0) = 29.30 \pm 0.28,\, \gamma = 2.19 \pm 0.99,\, \log_{10}(\beta_0) = 39.40 \pm 0.08,\,\delta = 1.02 \pm 0.22$. We assume a hardness ratio $\gamma = 1.8$ to convert between $2-10$\,keV and  $0.5-7$\,keV luminosities. 

\begin{figure*}
\centering
	\includegraphics[width=2.09\columnwidth]{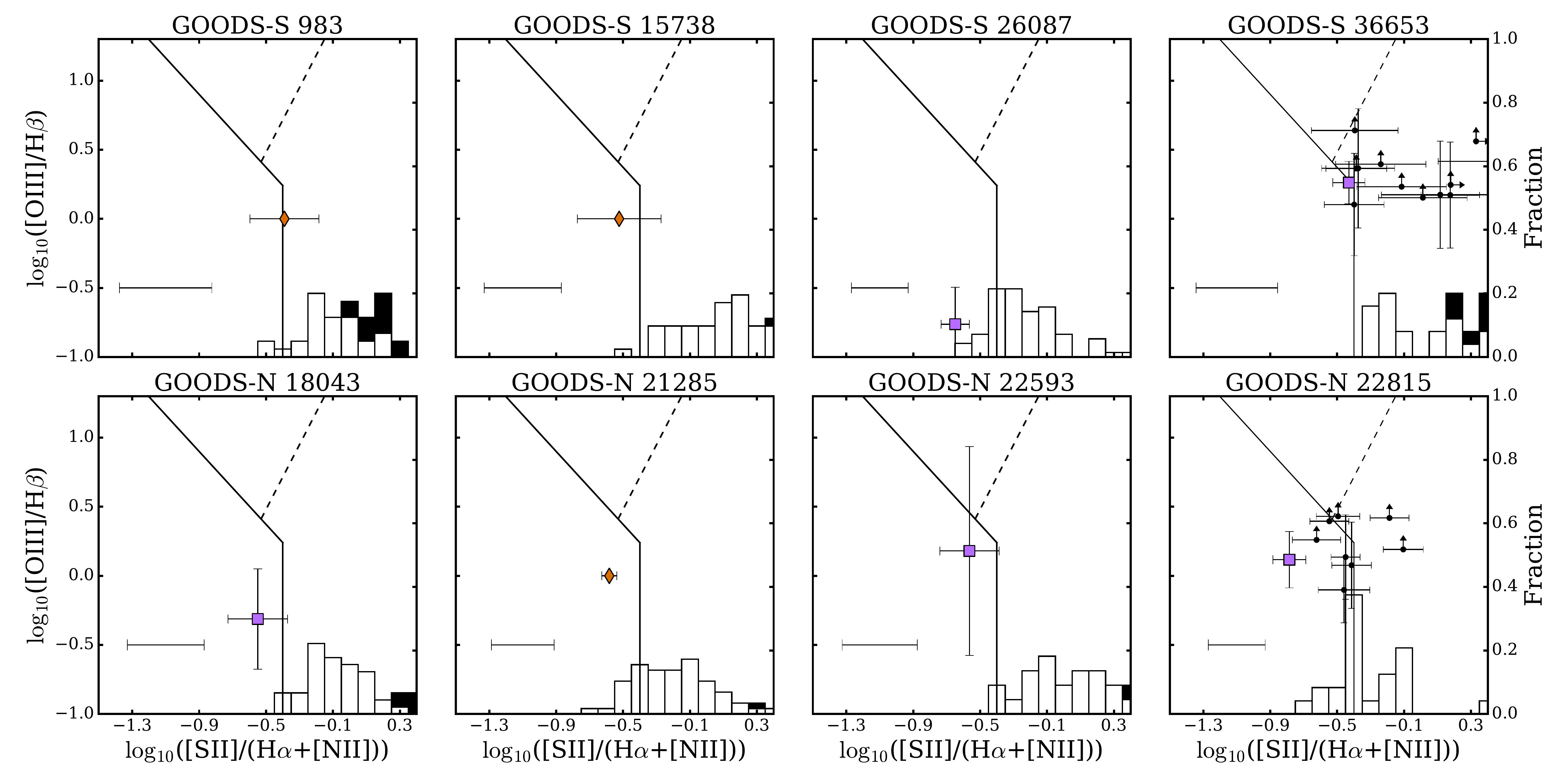}
	\caption{Modified BPT diagnostic for each LIER candidate. A fractional histogram of the [SII]/(H$\alpha$ + [NII]) emission is shown along the $x-$axis with median errorbars shown in the lower left. Black bars correspond to lower limits. Only GOODS-S 36653 and GOODS-N 22815 have detections in [OIII] and H$\beta$ and those pixels with both modified BPT ratios are plotted along with their uncertainties or limits. The total \texttt{Grizli} extracted line ratios are shown as colored shapes with corresponding errorbars. Orange diamonds have either no detected [OIII] or H$\beta$ while purple squares have all four lines detected.}
	\label{fig:BPTpaperH}  
\end{figure*}

\section{LIER Candidates}
\label{sec:LIER}

In this section we examine in detail the eight LIER candidates based on their morphologies, H$\alpha$ + [NII] emission line maps, emission line ratio maps and distribution of pixels in the modified BPT. 

We show color images of our candidates in Figure \ref{fig:RGBpaper}. All objects appear to have a red core surrounded by blue spiral arms indicating enhanced star formation. Figure \ref{fig:HaMAPpaper} shows the spatially resolved H$\alpha$ + [NII] emission line maps produced by the \texttt{Grizli} pipeline, highlighting the areas of strong emission in the galaxy in addition to showcasing the well resolved H$\alpha$ emission. For all objects, the emission line maps are extended and they usually trace the blue spiral arms. Finally, in Figures \ref{fig:LINERpaper} and \ref{fig:BPTpaperH}, we present the diagnostic emission line figures used to verify LIER candidacy for all eight objects. Emission line ratio maps for the  H$\alpha$+[NII] are presented in Figure \ref{fig:LINERpaper}. For a pixel's emission ratio to be plotted, we require a 2$\sigma$ detection in [SII] emission. However, if the H$\alpha$+[NII] flux is less than a 1$\sigma$ detection, we plot the lower limit of the emission ratio instead.  Ellipses (green) indicate 3, 6 and 9 kpc at the galaxy redshift using the ellipticity and position angle from \citet{vanderWel2014StrucParams} generated using stacked F125W, F140W, and F160W mosaics. The 1$\sigma$ \textit{Chandra} positions are marked in red. 

In Figure \ref{fig:BPTpaperH}, we plot the emission line ratios on our modified BPT diagram for each object along with the integrated emission line ratio. For a pixel's emission ratio to be plotted, we require a 2$\sigma$ detection in [SII] and [OIII] emission. However, if the corresponding H$\alpha$+[NII] or H$\beta$ flux is less than a 1$\sigma$ detection, we plot the lower limit of the emission ratio instead. Three of our objects have no available [OIII]/H$\beta$ ratio at all, (GOODS-S 983, GOODS-S 15738, and GOODS-N 21285) due to a non-detection in one or both lines. Another three (GOODS-S 26087, GOODS-N 18043, and GOODS-N 22593) have no individual pixels that satisfy the [OIII]/H$\beta$ SNR requirement as described above for both modified BPT emission line ratios. For these, we plot the [SII]/H$\alpha$+[NII] detections as a histogram along the $x$-axis. Only two of our objects (GOODS-S 36653 and GOODS-N 22815) have [OIII]/H$\beta$ detections per pixel that satisfy our SNR requirements. Each pixel with detections for both ratios for these objects are shown. Given that the [OIII]/H$\beta$ ratio is a strong indicator of AGN activity, its low values or non-detection in most of the objects in our sample is an indicator for the non-AGN origin of the LINER-like emission. 

\begin{itemize}
\item {\bf GOODS-S 983} presents significant extended LINER-like emission out to 6 kpc (Figure \ref{fig:LINERpaper}), mostly centered around the red core visible in Figure \ref{fig:RGBpaper}. This reinforces the notion that an evolved stellar population is providing the source of the ionization. Most of the pixels satisfy our criterion of [SII]/(H$\alpha$+[NII])$>-0.4$, placing them in the AGN regime of the modified BPT diagram. However, given the lack of a detection in [OIII]/H$\beta$ or a matched \textit{Chandra} source, it is unlikely that an active super-massive black hole is the source of LINER-like emission. 

\item {\bf GOODS-S 15738} exhibits strong diffuse emission out to 3 kpc in addition to a strand that extends out to 9 kpc (Figure \ref{fig:LINERpaper}). The latter emission appears to track strongly with a blue knot (Figure \ref{fig:RGBpaper}) which may indicate that this specific emission comes from a region of active star formation in line with the `starburst hypothesis' of extended emission. The large spatial extent of the emission, most of which satisfies the [SII]/(H$\alpha$+[NII]) criterion (Figure \ref{fig:BPTpaperH}), along with the lack of an [OIII]/H$\beta$ ratio detection suggest a non-AGN origin for the LINER-like emission. In addition, the corresponding \textit{Chandra} detection for the object satisfies none of the \citet{Luo17CDFS} AGN criteria and can be accounted for by black hole binary emission alone, further supporting the non-AGN hypothesis for ionization.

\item {\bf GOODS-S 26087} presents robust detections out to $\sim$6 kpc in almost every direction, most of which satisfies the requirement of [SII]/H$\alpha$+[NII]$>-0.4$ (\ref{fig:BPTpaperH}). Despite no corresponding $R$ or $K_s$ band information, the pure X-ray AGN criteria are not satisfied, and the X-ray emission can be accounted for by black hole binaries. Coupled with a low [OIII]/H$\beta$ ratio, the areas of LINER-like emission are likely not excited due to AGN activity. 

\item {\bf GOODS-S 36653} presents a region of LINER-like emission up to $\sim$6 kpc in the northeast portion of the galaxy (Figure \ref{fig:LINERpaper}). This corresponds to the edge of the visible core in Figure \ref{fig:RGBpaper}, an area that may be dominated by evolved stars. For this object, there are corresponding [OIII] to H$\beta$ detections which place all pixels confidently within the LINER regime. Furthermore, the overall \texttt{Grizli} emission measurements place the object as a whole in the LINER region of the modified BPT diagram, presenting an object that would likely be classified as an AGN in a typical spectroscopic survey. The discrepancy between the \texttt{Chandra} X-ray luminosity and estimated black hole binary contribution, a factor of about 2.3, is likely due to a contribution from hot gas, or large uncertainties in the SED modeled stellar mass. Like all the other candidates, GOODS-S 36653 does not satisfy any \textit{Chandra} X-ray criteria, again reinforcing the non-AGN origin for the ionization. 

\item {\bf GOODS-N 18043} shows extended LINER-like emission in the core along with a strip along the southeast spiral arm (Figures \ref{fig:RGBpaper} and \ref{fig:LINERpaper}). In addition, most of the pixels satisfy our [SII]/H$\alpha$+[NII] AGN selection (Figure \ref{fig:BPTpaperH}). While the integrated [OIII]/H$\beta$ ratio can be measured, there are no individual pixels with sufficient SNR for detection. The weak [OIII]/H$\beta$, coupled with the lack of a matched \textit{Chandra} detection point towards a non-AGN origin for LINER-like emission, whether from the red core, or from the bluer spiral arm. 

\item {\bf GOODS-N 21285} shows large spatial extent of LINER [SII]/(H$\alpha$+[NII]) ratios past the core, out to $\gtrsim$6 kpc. This emission tracks with the bluer regions seen in Figure \ref{fig:RGBpaper}, but  is also present in the redder core. Most of the individual pixels satisfying our SNR cuts do satisfy our [SII]/(H$\alpha$+[NII]) AGN selection threshold (Figure \ref{fig:BPTpaperH}). GOODS-N 21285 is our only candidate with matched observations across every band used in the \citet{Luo17CDFS} \textit{Chandra} X-ray criteria. Again, we find no evidence to suggest that the object is an AGN based on these detections and has X-ray emission consistent with black hole binaries. 

\item {\bf GOODS-N 22593} shows the most compact emission of all of our sources, however it still extends out to $\sim$3 kpc. Furthermore, given the lack of an X-ray detection and emission line ratios which place almost all pixels on the modified BPT in an area dominated by LINERs (Figure \ref{fig:BPTpaperH}), it is likely that the ionization is not due to an active nucleus. 

\item {\bf GOODS-N 22815} presents emission consistent with LINER-like ratios out to $\gtrsim$4 kpc. This is the second object, along with GOODS-S 36653, where we have measurements of the [OIII]/H$\beta$ ratios for individual pixels (Figure \ref{fig:BPTpaperH}). While the majority of the detected pixels lie below the [SII]/(H$\alpha$+[NII]) cut, the diffuse nature of the emission and the low [OIII]/H$\beta$ measurements do not support an active nucleus as the source of the ionizing radiation.  The discrepancy between the measured X-ray luminosity and the estimated galactic contribution, a factor of about 2.5, is likely due to a contribution from hot gas, or large uncertainties in the SED modeled stellar mass. Furthermore, the matching \textit{Chandra} observations do not satisfy any AGN criteria by themselves or when compared to observations in other bands.  
\end{itemize}

\section{Summary and Conclusions}
\label{sec:Conclusions}

In this paper, we have presented the identification of eight candidates for high redshift, $z\sim 0.9$, LIERs through spatially-resolved emission line diagnostics with \textit{Hubble}, and supported by corresponding \textit{Chandra} X-ray measurements and multi-wavelength photometry. All eight of our candidates present ionized regions consistent with LINER-like emission lines which are spatially extended, and therefore, inconsistent with the purely nuclear understanding of an AGN. Additionally, we find either weak or no detections for [OIII]/H$\beta$ ratios in six (75\%) of our objects, an emission diagnostic which is usually a strong indicator for AGN activity. Those objects with spatially resolved [OIII]/H$\beta$ serve to place the objects confidently in the LINER regime on our modified BPT diagram. Furthermore, most objects have X-ray luminosities consistent with galactic black hole binary emission, and none of our objects satisfy the AGN X-ray selection criteria outlined in \citet{Luo17CDFS}, once again reinforcing the hypothesis for the non-AGN origin for the ionizing radiation.

In a few of our LIER candidates, we find that the extended emission tracks in mainly in either the redder (GOOD-S 983) or bluer parts (GOODS-S 15783) of the optical image, consistent with both the `stellar' and `starburst' hypothesis for the ionization sources, but in general most of our objects have extended emission that overlaps both areas of the optical image. While the relatively high rates of specific star formation may suggest the `starburst hypothesis', we are unable to conclusively constrain the origin of the ionizing source.

This work serves to provide the first evidence for the existence of LIERs out to high redshift. This detection is made possible by the advancement in spatially resolved spectroscopy, allowing for the detection of diffuse emission throughout the candidates. This work was made possible by the \texttt{Grizli} grism reduction package and its ability to create drizzled emission line maps from multiple grism observations with an arbitrary number of position angles. Furthermore, we developed a modified BPT diagram, allowing for the separation between star-formation and AGN emission of objects where the [NII]/H$\alpha$ diagnostic is not available due to low spectral resolution. Despite the conservative nature of our cuts in our diagram, we are still able to find the clear examples of high redshift LIERs in the GOODSs fields. 

There are several projects that follow naturally from our work. Given that we cannot distinguish the ionizing source, it would be of interest to follow up these candidates to not only confirm the elevated line ratios but to gather diagnostics to test the various hypotheses of extended LINER emission. This could be attempted with higher resolution grism spectroscopy with the \textit{James Webb Space Telescope} (\textit{JWST}). It would be of interest to perform follow up X-ray observations of the objects, especially those in the GOODS North field which have shallower \textit{Chandra} observations. Observations would be well suited to current observational facilities, such as the \textit{Nuclear Spectroscopic Telescope Array} (\textit{NuSTAR}) or upcoming missions, such as the European Space Agency space observatory, Advanced Telescope for High-ENergy Astrophysics \citep[ATHENA]{barret13athena}, or the proposed NASA space observatory, \textit{Lynx} \citep{Weisskopf-Surveyor}.

Finally, the methods used in this work will also be applicable in the era of \textit{JWST}. \textit{JWST} is equipped with grism capabilities comparable to those of the \textit{Hubble Space Telescope} and the data analysis will benefit from employing the techniques outlined in this paper in the search for higher redshift LIERs. Slitless grism observations over large fields, which can be analyzed using \texttt{Grizli}, will serve as excellent methods to search for objects with follow-up potential using higher resolution spectroscopy, while also providing details about the total LIER population in the high redshift universe. 

\acknowledgments We would like to thank the anonymous referee for their constructive comments which improved the final paper. This work is supported in part by the Dartmouth E.E. Just program and the Space Telescope Science Institute. 

DM and KVN acknowledge support from program number HST-AR-14553, provided by NASA through a grant from the Space Telescope Science Institute, which is operated by the Association of Universities for Research in Astronomy, Incorporated, under NASA contract NAS5-26555.

This work is based on observations made with the NASA/ESA Hubble Space Telescope for 3D-HST Treasury Program (GO-12177 and GO-12328), GO-11359, GO-12190, GO-13420, GO-13779, GO-14227 as well as on support from program AR-14095 and AR-15553. The NASA/ESA Hubble Space Telescope is operated by the Association of Universities for Research in Astronomy, Inc., under NASA contract NAS5-26555.

The scientific results reported in this article are based in part on observations made by the \textit{Chandra} X-ray Observatory.

This data makes use of data from the Sloan Digital Sky Survey. Funding for SDSS-III has been provided by the Alfred P. Sloan Foundation, the Participating Institutions, the National Science Foundation, and the U.S. Department of Energy Office of Science. The SDSS-III web site is http://www.sdss3.org/.

SDSS-III is managed by the Astrophysical Research Consortium for the Participating Institutions of the SDSS-III Collaboration including the University of Arizona, the Brazilian Participation Group, Brookhaven National Laboratory, Carnegie Mellon University, University of Florida, the French Participation Group, the German Participation Group, Harvard University, the Instituto de Astrofisica de Canarias, the Michigan State/Notre Dame/JINA Participation Group, Johns Hopkins University, Lawrence Berkeley National Laboratory, Max Planck Institute for Astrophysics, Max Planck Institute for Extraterrestrial Physics, New Mexico State University, New York University, Ohio State University, Pennsylvania State University, University of Portsmouth, Princeton University, the Spanish Participation Group, University of Tokyo, University of Utah, Vanderbilt University, University of Virginia, University of Washington, and Yale University.

\bibliography{references}

% \appendix
% \section{appendix section}

\end{document}